\documentclass[prd,aps,showpacs,epsf,floats]{revtex4}%
\usepackage{amssymb}
\usepackage{amsfonts}
\usepackage{amsmath}
\usepackage{graphicx}
\usepackage{multirow}
\usepackage{color}%
\providecommand{\U}[1]{\protect\rule{.1in}{.1in}}

\begin{document}
\title{\textbf{CAUSATION ENTROPY FROM SYMBOLIC REPRESENTATIONS OF DYNAMICAL SYSTEMS}}
\author{Carlo Cafaro, Warren M. Lord, Jie Sun, and Erik M. Bollt}
\affiliation{Department of Mathematics, Clarkson University, 8 Clarkson Ave, Potsdam, NY,
13699-5815, USA}

\begin{abstract}
Identification of causal structures and quantification of direct information
flows in complex systems is a challenging yet important task, with practical
applications in many fields. Data generated by dynamical processes or
large-scale systems are often symbolized, either because of the finite
resolution of the measurement apparatus, or because of the need of statistical
estimation. By algorithmic application of causation entropy, we investigated
the effects of symbolization on important concepts such as Markov order and
causal structure of the tent map. We uncovered that these quantities depend
nonmontonically and, most of all, sensitively on the choice of symbolization.
Indeed, we show that Markov order and causal structure do not necessarily
converge to their original analog counterparts as the resolution of the
partitioning becomes finer.

\end{abstract}

\maketitle

\textbf{\textcolor{black}{While quantitative description and understanding of natural phenomena is at the core of science, inference of cause-and-effect relationships from measured data is a central problem in the study of complex systems, with many important practical applications. For example, knowing ``what causes what" allows for the effective identification of the cause of a medical disease or disorder, and for the detection of the root source of defects of engineering systems.
However, the act of measuring the states of a dynamical system mediates the inference of cause-and-effect relationships. For instance, all observation procedures carry the limitation of finite precision. A common example is the binning of data (histograms) into discrete symbols.
In this paper we use a toy mathematical model to show that such digitization (symbolization) may lead to inferred causal relationships that differ significantly from those of the original system, even when the amount of data is unlimited.}}
\textbf{Although based on a simple mathematical model, our results shed new
light on the challenging nature of causality inference.}

\section{Introduction}

Uncovering cause-and-effect relationships remains an exciting challenge in
many fields of applied science. For instance, identifying the causes of a
disease in order to prescribe effective treatments is of primary importance in
medical diagnosis \cite{lesne}; locating the defects that could cause abrupt
changes of the connectivity structure and adversely affect the performance of
the system is a main objective in structural health monitoring
\cite{gao,vicente}. Consequently, the problem of inferring causal
relationships from observational data has attracted much attention in recent
years~\cite{lesne,gao,vicente,schreiber2000,rothman05,palus-report,frenzel07,guo08,heckman08,barrett10,cubitt12,runge12a,runge12b,marinazzo12,sun2014a,
sun2014b, sun2014c,porta14}.

Identifying causal relationships in large-scale complex systems turns out to
be a highly nontrivial task. As a matter of fact, a reliable test of causal
relationships requires the effective determination of whether the
cause-and-effect is real or is due to the secondary influence of other
variables in the system. This, in principle, can be achieved by testing the
relative independence between the potential cause and effect conditioned on
all other variables in the system. Such a method essentially demands the
estimation of joint probabilities for (very) high dimensional variables from
limited available data and suffers the curse of dimensionality. In practice,
there are various approaches in statistics and information theory that aim at
accomplishing the proper conditioning without the need of testing upon all
remaining variables of the system at once~\cite{palus-report,PCbook}.
The basic idea behind many such approaches originates from the classical
PC-algorithm~\cite{PCbook}, which repeatedly measures the relative
independence between the cause and effect conditioned on combinations of the
other variables. As an alternative, we recently developed a new entropy-based
computational approach that infers the causal structure via a two-stage
process, by first aggregatively discovering potential causal relationships and
then progressively removing those (from the stage) that are
redundant~\cite{sun2014a, sun2014b, sun2014c}.

In almost all computational approaches for inferring causal structure, it is
necessary to estimate the joint probabilities underlying the given process.
Large-scale data sets are commonly analyzed via discretization procedures, for
instance using binning, ranking, and/or permutation
methods~\cite{darbellay,bandt2002,amigo2005,staniek,kugi,haruna2013}. These
methods generally require fine-tuning of parameters and can be sensitive to
noise. On the other hand, the time-evolution of a physical system can only be
measured and recorded to a finite precision, resembling an approximation of
the true underlying process. This finite resolution can be characterized by
means of a finite set of symbols, yielding a discretization of the phase
space. Regardless of the nature and motivation of discretization, the precise
impacts on the causal structure of the system is essentially unexplored. Here,
we investigate the symbolic description of a dynamical system and how it
affects the resulting Markov order and causal structures.
Such description, based on partitioning the phase space of the system, is also
commonly known as symbolization. Symbolization converts the original dynamics
into a stochastic process supported on a finite sample space.
Focusing on the tent map for the simplicity, clarity and completeness of
computation it allows~\cite{tim}, we introduce numerical procedures to compute
the joint probabilities of the stochastic process resulting from arbitrary
partitioning of the phase space. Furthermore, we develop causation entropy, an
information-theoretic measure based on conditional mutual information as a
mean to determine the Markov order and (temporal) causal structure of such
processes. We uncovered that a partitioning that maintains dynamic invariants
of the system does not necessarily preserve its causal structure. On the other
hand, both the Markov order and causal structure depend nonmonotonically and,
indeed, sensitively on the partitioning.

\section{Phase Space Partitioning and Symbolic Dynamics}

A powerful method of analyzing nonlinear dynamical systems is to study their
symbolic dynamics through some topological partition of the phase
space~\cite{tim, erikbook, gora}. The main idea characterizing symbolic
dynamics is to represent the state of the system using symbols from a finite
alphabet defined by the partition, rather than using a continuous variable of
the original phase space. For more details, we refer to \cite{lind,
lind1,kitchens, robinson}. The issue of partitioning was shown to affect
entropic computations in a nontrivial manner~\cite{erik2000, erik2001} and, as
we will highlight in the paper, is also intricate and central to a general
information-theoretic description of the system.

\subsection{Partition of the phase space and symbolic dynamics}

\begin{figure}[ptb]
\centering
\includegraphics[width=0.39\textwidth]{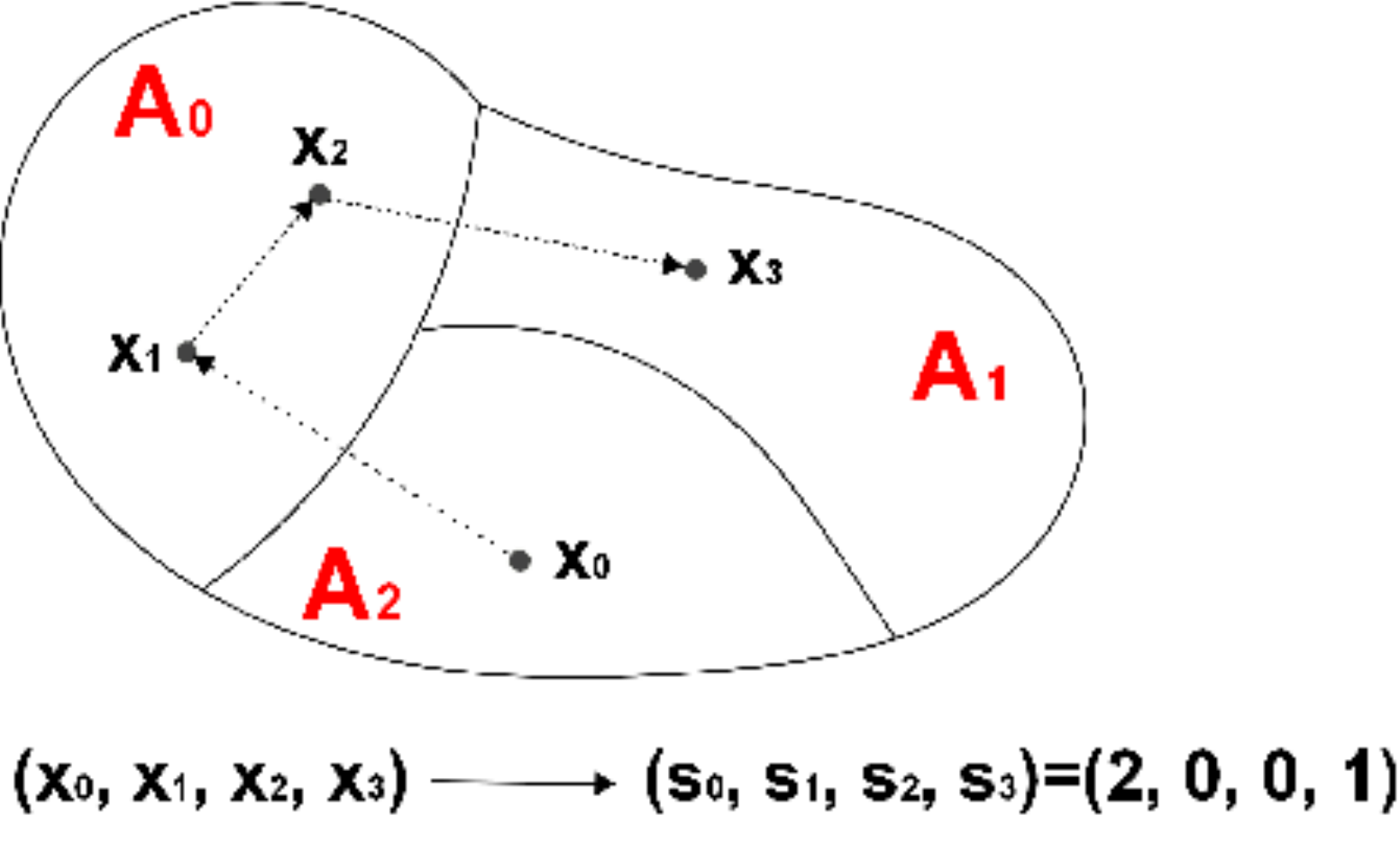}\caption{Schematic illustration of
partitioning the phase space and the resulting symbolic dynamics. Given the
partitioning $\mathcal{A}=\{A_{0},A_{1},A_{2}\}$, the trajectory $(x_{0}%
,x_{1},x_{2},x_{3},\dots)$ leads to a symbol sequence $(s_{0},s_{1}%
,s_{2},s_{3},\dots)=(2,0,0,1,\dots)$. }%
\label{fig1}%
\end{figure}

Consider a discrete dynamical system given by
\begin{equation}
x_{t+1}=f(x_{t}),
\end{equation}
where $x_{t}\in M\subset\mathbb{R}^{d}$ represents the state of the system at
time $t$ and the vector field $f:M\rightarrow M$ governs the dynamic evolution
of the states. A (\emph{topological}) \emph{partition} of the phase space $M$
is a finite collection $\mathcal{A}\overset{\text{def}}{=}\left\{
A_{0}\text{,..., }A_{m}\right\}  $ of disjoint open sets whose closures cover
$M$, i.e.,
\begin{equation}
M={\displaystyle\bigcup\limits_{i=0}^{m}}\bar{A}_{i}. \label{eq:partition}%
\end{equation}
The partition leads to the corresponding \textit{symbolic dynamics}. In
particular, for any trajectory $\{x_{0},x_{1},x_{2},\dots\}$ of the original
dynamics contained in the union of $A_{i}$'s, the partition yields a
\textit{symbol sequence} $\{s_{0},s_{1},s_{2},\dots\}$ given by
\begin{equation}
s_{t}=\sum_{i=0}^{m}\chi_{A_{i}}(x_{t})\cdot i,
\end{equation}
where $\chi_{A}$ is the indicator function defined as
\begin{equation}
\chi_{A}(x)=%
\begin{cases}
1,~\mbox{if~$x\in A$},\\
0,~\mbox{if~$x\notin A$}.
\end{cases}
\end{equation}
In other words, the symbolic state $s_{t}$ is determined by the open set
$A_{i}$ that contains the state $x_{t}$. See Fig.~\ref{fig1} for a schematic illustration.

In general, the same symbol sequence may result from distinct trajectories. If
the partition is \textit{generating}, then every symbol sequence corresponds
to a unique trajectory \cite{rudolph}. A special case is the so-called Markov
partition~\cite{erikbook, skufca}, for which the transition from one symbolic
state to another is independent of past states, analogous to a Markov process.
On the other hand, a generating partition is not necessarily
Markov~\cite{gora, collet}.

The precise effects of partitioning on the symbolic dynamics remains an
interesting and challenging problem, with recent progress in a few directions.
Focusing on the equivalence between the original and symbolic dynamics,
Bollt~\textit{et.~al.} studied the consequence of misplaced partitions on
dynamical invariants~\cite{erik2000, erik2001}, while Teramoto and Komatsuzaki
investigated topological change in the symbolic dynamics upon different
choices of Markov partitions~\cite{teramoto}. On the other hand, the degree of
self-sufficiency of the symbolic dynamics, irrespective of the equivalence to
the original dynamics, has started to gain increasing interest, focusing on
information-theoretical measures such as information closure and prediction
efficiency~\cite{pfante}. We here adopt a different perspective and study how
causal structures emerge and/or change under different choices of partitioning.

\subsection{From dynamical systems to stochastic processes via symbolic
dynamics}

The symbolic description of a dynamical system leads naturally to an
interpretation of such systems as stochastic processes \cite{gallager2013}.
Let $(M,\Sigma,\mu)$ be a measure space with Borel field $\Sigma$ and
probability measure $\mu$ such that $\mu:\Sigma\rightarrow\lbrack0,1]$ and
$\mu(M)=1$. Furthermore, assume that $\mu$ is the unique ergodic invariant
measure under the mapping $f$, that is
\begin{equation}%
\begin{cases}
\mbox{(Invariance)}~\mbox{For every
$B\in\Sigma$, $\mu(f^{-1}(B))=\mu(B)$.}\\
\mbox{(Ergodicity)}~\mbox{For
every $B\in\Sigma$ with $f^{-1}(B)=B$, either $\mu(B)=0$ or $\mu(B)=1$.}
\end{cases}
\label{eq:invariant}%
\end{equation}
Given the partitioning defined by Eq.~\eqref{eq:partition}, the symbol space
(alphabet) $\Omega$ is made of $m+1$ symbols (alphabet letters),
\begin{equation}
\Omega\overset{\text{def}}{=}\{0,1,\dots,m-1,m\}.
\end{equation}
We can formally define a random variable $S$ as a measurable function
$S:\Omega\rightarrow\mathbb{R}$ with the probabilities given by
\begin{equation}
P(s_{t})\overset{\text{def}}{=}\mbox{Prob}(S_{t}=s_{t})=\mu(A_{s_{t}%
}),~\forall{s_{t}\in\Omega}. \label{eq:prob}%
\end{equation}
This line of reasoning can be generalized to accommodate joint probabilities
of arbitrary finite length,
\begin{align}
P(s_{t},s_{t+1},s_{t+2},\dots)  &  \overset{\text{def}}{=}\mbox{Prob}(S_{t}%
=s_{t},S_{t+1}=s_{t+1},S_{t+2}=s_{t+2},\dots)\nonumber\label{eq:prob2}\\
&  =\mu(A_{s_{t}}\cap f^{-1}(A_{s_{t+1}})\cap f^{-2}(A_{s_{t+2}})\dots).
\end{align}
The probabilities in Eqs.~\eqref{eq:prob} and~\eqref{eq:prob2} are
time-invariant because $\mu$ is invariant as assumed in
Eq.~\eqref{eq:invariant}. Within this setting, $P(s)$ denotes the probability
that the symbolic state of the system (at any time) is equal $s$, while
$P(s,s^{\prime})$ is the probability of the current and next symbols being $s$
and $s^{\prime}$, respectively. Therefore, this framework defines a discrete
stochastic process where the symbolic states are regarded as random variables
whose stationary joint distributions are determined by Eqs.~\eqref{eq:prob}
and~\eqref{eq:prob2}. We point out that the support of such a stochastic
process associated with the symbolic dynamics\textbf{\ }is commonly referred
to as a shift space~\cite{erikbook}.

\section{Markov Order, Causal Structure, and Inference}

For a given symbolization of a dynamical system that originates from a chosen
partitioning of the phase space, we are interested in defining and identifying
a minimal set of past states that encode information about the current state
$S_{t}$. This will enable us to remove redundant information of the past when
making efficient predictions about the future.

\subsection{Markov Order and Causal Structure of the Symbolic Dynamics}

\begin{figure}[ptb]
\centering
\includegraphics[width=0.5\textwidth]{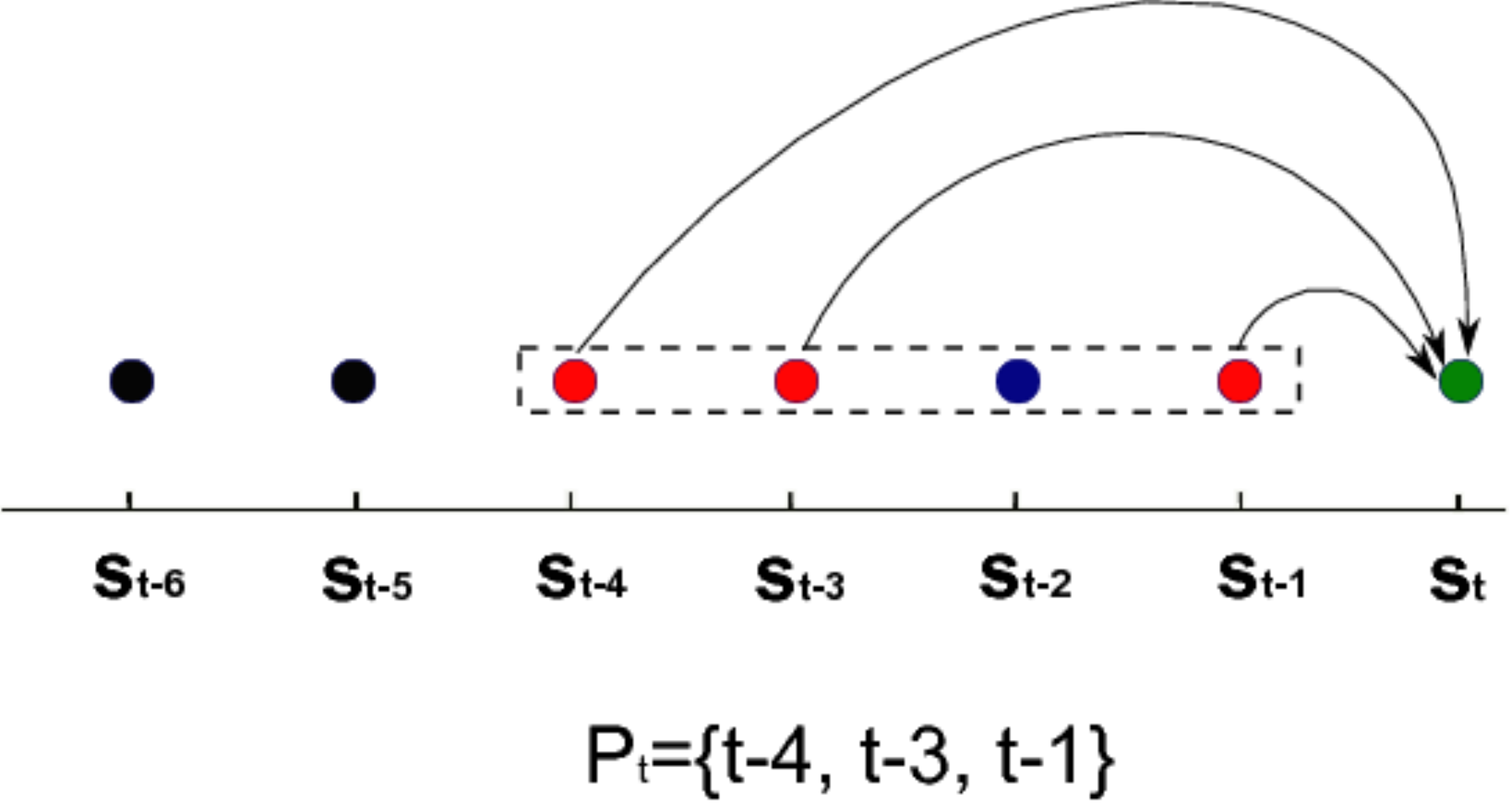}\caption{(Color online) An example
causal structure of a Markov process. Here the process is of order $k=4$,
although only three (marked in red) out of the four past time indices
(enclosed by dashed box) are needed to render the current state (green)
conditionally independent of the rest of the past. The set of causal time
parents of $t$ is therefore $\mathcal{P}_{t}=\{t-4,t-3,t-1\}$ in
Eq.~\eqref{eq:causal}. }%
\label{fig2}%
\end{figure}

In view of the probabilistic interpretation of the symbol dynamics, we refer
to a partition as Markovian of order $k$ if the resulting stochastic process
is Markov order $k$; that is, if the symbolic state only depends on its past
$k$ states rather than on the entire history. Using the following notations
\begin{equation}%
\begin{cases}
s_{t^{-}}\overset{\text{def}}{=}(s_{t-1},s_{t-2},\dots),\\
s_{t^{-}}\backslash s_{(t-k)^{-}}\overset{\text{def}}{=}(s_{t-1},s_{t-2}%
,\dots,s_{t-k}),
\end{cases}
\end{equation}
a process is Markov order $k$ if and only if the conditional probabilities
satisfy
\begin{equation}
P(s_{t}|s_{t-})=P(s_{t}|s_{t^{-}}\backslash s_{(t-k)^{-}}) \label{eq:Markov}%
\end{equation}
for every choice of $s_{t^{-}}$ and no nonnegative integer smaller than $k$
fulfills this requirement. (When $k=0$, we call the process an
i.i.d.\textbf{\ }process.) In other words, information carried in the past
states\textbf{\ }is all conditionally redundant given information about the
past $k$ states. On the other hand, there might be further redundancy in the
information encoded in these $k$ states. In particular, let
\begin{equation}
\mathcal{P}_{t}\subset\{t-k,t-k+1,\dots,t-1\}
\end{equation}
be a minimal set contained in the Markov time indices for which
\begin{equation}
P(s_{t}|s_{t-})=P(s_{t}|s_{\mathcal{P}_{t}})\label{eq:causal}\\
\end{equation}
holds for every $s_{t^{-}}$. Therefore for every proper subset $\mathcal{P}%
_{t}^{\prime}$ of $\mathcal{P}_{t}$ Eq. (\ref{eq:causal}) does not hold. We
refer to $\mathcal{P}_{t}$ as the set of \textit{causal time parents} of time
$t$. Conditioning on the states with time indices given by $\mathcal{P}_{t}$,
information of all other states becomes redundant. The states at time(s)
$\mathcal{P}_{t}$ are the only ones that cause the current state, and
therefore the set $\mathcal{P}_{t}$ defines a causal structure of the symbolic
dynamics. This can be viewed as a finer description than the Markov order,
which in turn allows for a more efficient encoding of the process.
Figure~\ref{fig2} illustrates the difference between Markov order and causal
structure of an example process.

\subsection{Entropy, Mutual Information, Conditional Mutual Information, and
Causation Entropy}

Practical evaluation of joint probabilities is delicate for two reasons:
first, numerical imperfections due to finite precision of the computing
machines are unavoidable; second, when the probabilities need to be estimated
from finite data samples, estimation errors are inevitable. Naturally, the
appearance of such numerical and estimation errors will propagate into
Eqs.~\eqref{eq:Markov} and~\eqref{eq:causal}, making it difficult to
distinguish equalities from inequalities. These equations need to be examined
for joint sequences, leading to an overwhelming number of (heuristic)
decisions that need to be made. This, in turn, renders unreliable the direct
determination of Markov order and causal structure based on their respective
formal definitions. From a statistical standpoint, it is preferable to base
such determination on a minimal number of equations/decisions.
Appropriately defined information-theoretic measures fulfill this goal by
collectively grouping the joint probabilities, therefore greatly reducing the
number of equations/decisions.

Recall that Shannon entropy is a quantitative measure of the uncertainty of a
random variable. For a discrete random variable $X$ with probability mass
function $P(x)\overset{\text{def}}{=}\mbox{Prob}(X=x)$, its entropy is defined
as~\cite{shannon}
\begin{equation}
H(X)=-\sum_{x}P(x)\log{P(x)},
\end{equation}
where $\log$ is taken to be base $2$ throughout the paper. The mutual
information between two random variables $X$ and $Y$ is given by~\cite{cover}
\begin{equation}
I(X;Y)=\sum_{x,y}P(x,y)\log\frac{P(x,y)}{P(x)P(y)}.
\end{equation}
Mutual information measures the deviation from independence between $X$ and
$Y$. It is generally nonnegative and equals zero if and only if $X$ and $Y$
are independent. Similarly, the conditional mutual information between $X$ and
$Y$ given $Z$ is defined as~\cite{cover}
\begin{equation}
I(X;Y|Z)=\sum_{x,y,z}P(x,y,z)\log\frac{P(x,y|z)}{P(x|z)P(y|z)},
\end{equation}
and it measures the reduction of uncertainty of $X$ ($Y$) due to $Y$ ($X$)
given $Z$. Conditional mutual information is nonnegative, and equals zero if
and only if $X$ and $Y$ are conditionally independent given $Z$.

For a stationary stochastic process $\{S_{t}\}$ and a given set of time
indices $J_{t}\subset t^{-}$, we propose to define the (temporal) causation
entropy (CSE) from $J_{t}$ to $t$ to be
\begin{equation}
C_{J_{t}\rightarrow t}=I(S_{J_{t}};S_{t}|S_{t^{-}\backslash J_{t}}).
\label{newdef}%
\end{equation}
Being a conditional mutual information, causation entropy is always
nonnegative. It is strictly positive if and only if uncertainty about the
state $S_{t}$ is reduced due to the knowledge about $S_{J_{t}}$. This occurs
when the past states with time indices $J_{t}$ carry information about the
current state at time $t$. We remark that Eq. (\ref{newdef}) is an adapted
definition of causation beyond our previous work~\cite{sun2014a, sun2014b,
sun2014c} \ for a specific scenario, in the sense that direct causality is now
intimately linked to causation entropy being strictly positive without the
need of appropriately choosing the conditioning set.

\subsection{Inference of Markov Order and Causal Structure}

Based on the definition of Markov in Eq.~\eqref{eq:Markov}, a stochastic
process has Markov order $k$ if and only if
\begin{equation}
C_{(t-k)^{-}\rightarrow t}=0 \label{eq:CSEmarkov}%
\end{equation}
for the smallest possible nonnegative integer $k$. Algorithmically, we start
by examining Eq.~\eqref{eq:CSEmarkov} for $k=0$. If it holds true, then the
process is i.i.d. If not, we proceed with $k=1,2,\dots$, until the equation is
satisfied. The resulting value of $k$ is the Markov order of the process. As a
side remark, we note that there are other entropy-based approaches to
determine the Markov order~\cite{ragwitz2002,papapetrou,pethel2014}.

Now we discuss the inference of causal structure via causation entropy. Given
the definition of causal structure in Eq.~\eqref{eq:causal}, it follows that a
Markov process of order $k$ has causal time parents $\mathcal{P}_{t}$ if and
only if
\begin{equation}
C_{(t^{-}\backslash\mathcal{P}_{t})\rightarrow t}=0 \label{eq:CSEcausal}%
\end{equation}
where $\mathcal{P}_{t}\subset\{t-k,t-k+1,\dots,t-1\}$ and no proper subset of
$\mathcal{P}_{t}$ fulfills the equation. Computationally, it is generally
infeasible to efficiently find the causal structure without additional
assumptions about the underlying joint distributions. A general assumption,
called the faithfulness or stability assumption, requires that the joint
effect/cause is decomposable into individual components
~\cite{sun2014b,PCbook,pearl}. That is to say, for every $t^{\prime}%
\in\mathcal{P}_{t}$, the contribution measured in terms of the conditional
mutual information $I(S_{t^{\prime}};S_{t}|S_{\mathcal{Q}_{t}})$ is
non-vanishing for every $\mathcal{Q}_{t}$ that does not include $t^{\prime}$
or $t$. Under this assumption, we can show that the causal time parents
$\mathcal{P}_{t}$ form the minimal set of time indices that maximizes
causation entropy~\cite{sun2014b}, i.e.,
\begin{equation}
\label{oCSE}\mathcal{P}_{t}=\underset{J_{t}\subset\mathcal{K}}\bigcap
J_{t},~\mbox{where
$\mathcal{K}=\{J\subset t^-:\forall K\subset t^-, C_{J\rightarrow t}\geq
C_{K\rightarrow t}\}$}.
\end{equation}
We refer to Eq. (\ref{oCSE}) to as the optimal causation entropy principle,
which allows the transformation of the causal inference problem into a
numerical optimization problem.

Algorithmically, we propose to infer the causal set $\mathcal{P}_{t}$ via a
two-stage iterative process, described as follows. The first stage, which we
call aggregative discovery, starts by finding a time index $t^{\prime-}$ which
maximizes the mutual information $I(S_{t^{\prime}};S_{t})$ provided that such
mutual information is strictly positive. That is,
\begin{equation}
\label{eq:ocsef1}t_{1}=\underset{t^{\prime-}}{\operatorname{argmax}%
}~I(S_{t^{\prime}};S_{t}).
\end{equation}
Then, at each subsequent step, a new time index $t_{l+1}$ is identified among
the rest of the indices to maximize the conditional mutual information given
the previously selected time indices, that is,
\begin{equation}
\label{eq:ocsef2}t_{l+1}=\underset{t^{\prime}\in\left(  t^{-}\backslash
\{t_{1},t_{2},\dots,t_{l}\}\right)  } {\operatorname{argmax}}~I(S_{t^{\prime}%
};S_{t}|S_{t_{1},t_{2},\dots,t_{l}}).
\end{equation}
Such iterative process ends when the corresponding maximum conditional mutual
information equals zero, and the outcome yields a set of time indices
$\mathcal{Q}_{t}=\{t_{1},t_{2},\dots,t_{L}\}\supset\mathcal{P}_{t}$.

Then, in the second stage, we progressively remove time indices in
$\mathcal{Q}_{t}$ that are redundant (i.e., do not belong to $\mathcal{P}_{t}%
$). In particular, we enumerate through the time indices in $\mathcal{Q}_{t}$
and remove each component $t_{l}$ for which
\begin{equation}
I(S_{t_{l}};S_{t}|S_{\mathcal{Q}_{t}\backslash t_{l}})=0.
\end{equation}
Every time a component is removed, the set $\mathcal{Q}_{t}$ is updated
accordingly. The end of the process is then inferred as the set of causal time
parents $\mathcal{P}_{t}$. We remark that the discovery and removal stages of
our algorithm are reminiscent of the forward selection and backward
elimination in regression analysis~\cite{Draper}. Here, for the purpose of
correct and consistent inference of Markov order and causal structure, we have
adopted conditional mutual information in our algorithm.~\cite{sun2014b}.

Two practical considerations need to be taken into account for the inference
of Markov order and causal structure. First, the history of a variable needs
to be truncated, i.e., $t^{-}$ will be approximated by $t^{-}\approx
\{t-T,t-T+1,\dots,t-1\}$ for some $T\gg1$ in Eq.~\eqref{eq:CSEmarkov}
(regarding Markov order) and Eqs.~(\ref{eq:ocsef1}-\ref{eq:ocsef2}) (regarding
causal structure). In particular, such truncation leads to a partial
fulfillment of both the Markov requirement in Eq.~\eqref{eq:Markov} and causal
structure in Eq.~\eqref{eq:causal}. Second, numerical and estimation errors
generally render information-theoretic quantities such as the mutual
information, conditional mutual information, and causation entropy nonzero
(and in particular, even negative~\cite{grassberger1988,grassberger2008}). In
order to decide whether or not an estimate should be regarded as zero, one
needs a threshold-selecting procedure~\cite{sun2014b}: an estimated quantity
smaller than a predefined threshold will be considered vanishing.


\section{Markov Order and Causal Structure from the Symbolization of Tent Map}

In this section we provide an application of our theoretical procedure in
determining the Markov order and causal structure of symbolic dynamics of the
tent map. The primary reason why we have chosen the one-dimensional tent map
as an example is twofold. First, the tent map is simple enough to allow
explicit analytical computations of the entropic functionals of known
probability distributions. Such computations are not only useful for
cross-checking numerical estimates, they also provide some insights into the
information-theoretic measures employed in our investigation. Second,
regardless of its simple form, the tent map appears to serve as a rich
test-bed for the investigation of how Markov order and causal structure of a
dynamical system are affected by the choice of symbolization. In fact, under
symbolization, even a 1D map such as the tent map can be regarded quite
complex from a topological
standpoint~\cite{erikbook,gora,lind,lind1,kitchens,robinson}. Finally, we
remark that our computational framework can be applied to arbitrary unimodal maps.

\subsection{Tent Map and Partitioning}

The tent map is a one-dimensional system given by $x_{t+1}=T(x)$ where
$T:[0,1]\rightarrow\lbrack0,1]$ is defined as
\begin{equation}
T(x)\overset{\text{def}}{=}%
\begin{cases}
2x, & \mbox{if~$0\leq x\leq \frac{1}{2}$},\\
2(1-x), & \mbox{if~$\frac{1}{2}<x\leq1$}.
\end{cases}
\end{equation}
Specifically, we shall discuss the manner in which different choices of the
partitioning lead to (qualitatively and quantitatively) different
symbolizations of the original dynamics with specific Markov orders and causal
structures. For the time being, we limit our investigation to a binary
symbolic description of the dynamical map. Consider a general binary
partitioning of the phase space defined by the parameter $\alpha\in(0,1)$, so
that
\begin{equation}
\mathcal{A}=\{A_{0},A_{1}\}=\{[0,\alpha),(\alpha,1]\}.
\end{equation}
Such partitioning allows us to represent a continuous trajectory by a sequence
of binary symbols (bits).
We remark that\textbf{\ }the choice of $\alpha=0.5$ leads to a generating
partition which gives rise to a symbolic dynamics that is topologically
equivalent to the original system~\cite{erik2000, erik2001, erikbook}.

\subsection{Invariant Probability Measure and Joint Probabilities}

\begin{figure}[ptb]
\centering
\includegraphics[width=0.59\textwidth]{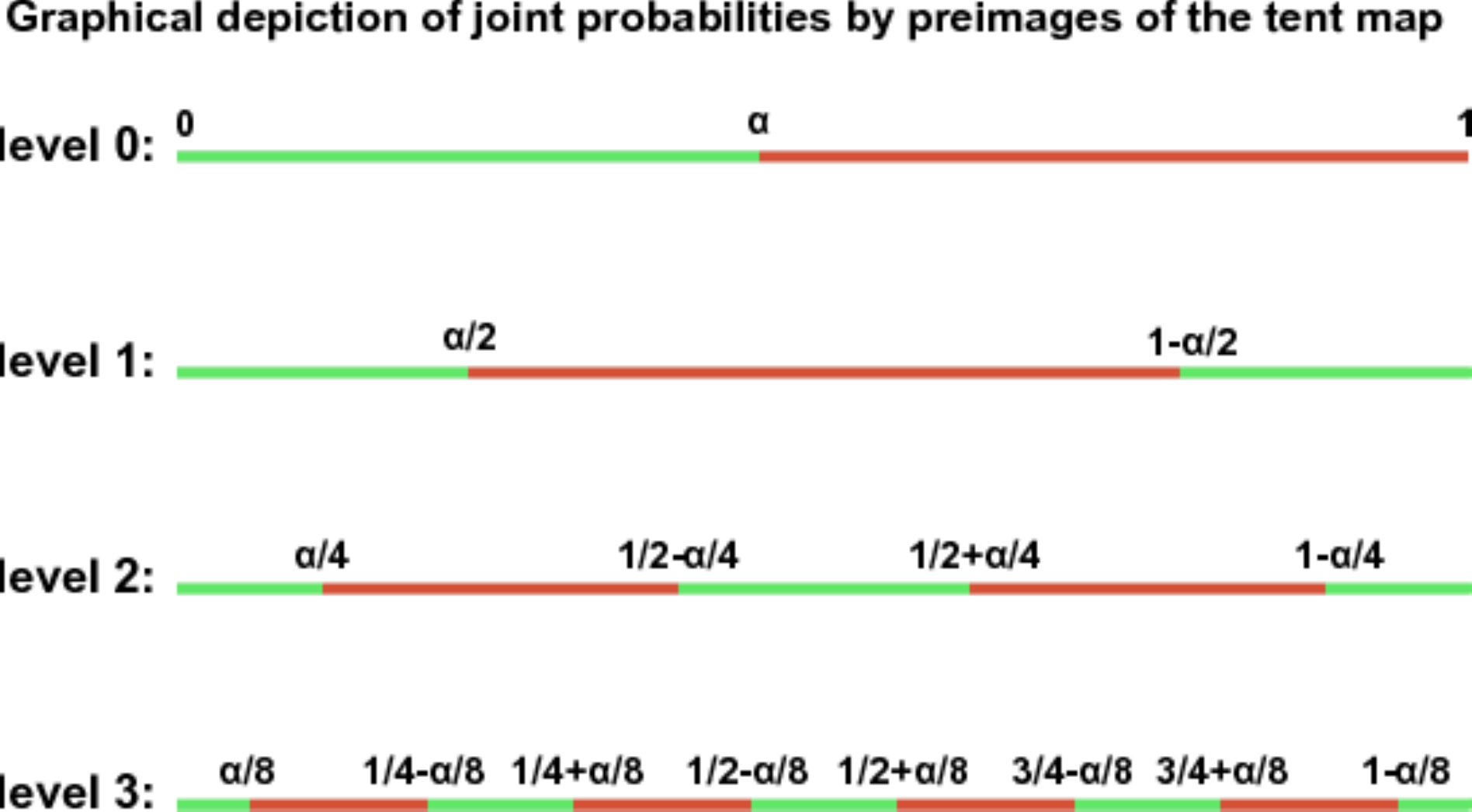}\caption{(Color online) Preimages
of partitioning intervals of the tent map. The intervals $I_{l}^{(0)}$ (green)
and $I_{l}^{(1)}$ (red) are defined by Eq.~\eqref{eq:preimage} and are shown
for levels $l=0,1,2,3$ for the choice of $\alpha=0.45$. In general, at each
level $l$, the subintervals start from $I_{l}^{(0)}$ and then alternate in
between $I_{l}^{(1)}$ and $I_{l}^{(0)}$. The relative ordering of the
subintervals across levels can change for different values of $\alpha$,
although they remain the same as shown in the picture for all $\alpha
\in(4/9,4/7)$.}%
\label{fig4}%
\end{figure}

The unique ergodic invariant measure of the tent map can be found by solving
the first equation in~\eqref{eq:invariant} (also called a continuity
equation)
for each subinterval of $[0,1]$, leading to
\begin{equation}
\mu([a,b])=b-a.
\end{equation}
This immediately gives $P(0)=\alpha$ and $P(1)=1-\alpha$. From
Eq.~\eqref{eq:prob2}, the joint probability of an arbitrary sequence of length
$n+1$ is determined by
\begin{equation}
P(s_{0},s_{1},s_{2},\dots,s_{n})=\mu\left(  \bigcap\limits_{l=0}^{n}%
I_{l}^{\left(  s_{l}\right)  }\right)  \text{,} \label{eq:tentprob}%
\end{equation}
where the intervals are defined by the preimages of $[0,\alpha)$ and
$(\alpha,1]$ as
\begin{equation}%
\begin{cases}
I_{l}^{\left(  0\right)  }\overset{\text{def}}{=}\left\{  x\in\left[
0,1\right]  :T^{l}\left(  x\right)  \in\left[  0\text{, }\alpha\right)
\right\}  ,\\
I_{l}^{\left(  1\right)  }\overset{\text{def}}{=}\left\{  x\in\left[
0,1\right]  :T^{l}\left(  x\right)  \in\left(  \alpha\text{, }1\right]
\right\}  .
\end{cases}
\label{eq:preimage}%
\end{equation}
In other words, the initial conditions corresponding to a specific symbolic
string of length $n$ are formed by a finite disjoint union of intervals.
Figure~\ref{fig4} shows an example of these intervals $\{I_{l}^{(0)}%
,I_{l}^{(1)}\}$ for $\alpha=0.45$ and four levels $l=0,1,2,3$.

This offers a computationally feasible description with which joint
probabilities can be calculated. From Eq.~\eqref{eq:tentprob}, we obtain that
for $n=0$, $P(s_{0})=\mu(I_{0}^{(s_{0})})$, giving $P(0)=\alpha$ and
$P(1)=1-\alpha$ as expected. For $n=1$, we have $P(s_{0},s_{1})=\mu
(I_{0}^{(s_{0})}\cap I_{1}^{(s_{1})})$. This gives the probabilities
$P(0,0)=\alpha/2$, $P(0,1)=\alpha/2$, $P(1,0)=1-3\alpha/2$, and $P(1,1)=\alpha
/2$ for all $\alpha<2/3$ (see also Fig.~\ref{fig4}). For general values of
$n$, we proceed as follows. First, we define the level-$l$ preimages of
$\alpha$ to be $\{\alpha_{l}^{(i)}\}$ ($i=1,2,\dots,2^{l}$), which are the
roots of the equation
\begin{equation}
T^{l}(x)-\alpha=0.
\end{equation}
For convenience, we sort $\{\alpha_{l}^{(i)}\}$ in the ascending order of $i$
and, additionally, define $\alpha_{l}^{(0)}\overset{\text{def}}{=}0$ and
$\alpha_{l}^{(2^{l}+1)}\overset{\text{def}}{=}1$. Then, the preimages sets of
$[0,\alpha)$ and $(\alpha,1]$ as introduced in Eq.~\eqref{eq:preimage} can be
explicitly computed as (for every $l\geq1$)
\begin{equation}%
\begin{cases}
I_{l}^{(0)}=\bigcup_{i=0}^{2^{l-1}}(\alpha_{l}^{(2i)},\alpha_{l}^{(2i+1)}),\\
I_{l}^{(1)}=\bigcup_{i=1}^{2^{l-1}}(\alpha_{l}^{(2i)-1},\alpha_{l}^{(2i)}).
\end{cases}
\end{equation}
Such preimages sets are subsequently used to calculate joint probabilities.
Note that for symbolic strings of length $n$, both the total number of joint
probabilities and the total number of intervals contributing to these
probabilities equal $2^{n}$.

Note that the joint probability of the symbol sequence $P(s_{0},s_{1}%
,s_{2},\dots,s_{n})$ depends on the particular choice of the partitioning
point $\alpha$. However, the functional $\alpha$-dependences of such
probabilities remain the same for all $\alpha$ values within intervals
determined by the $2^{n}$ distinct roots $\{x_{\ast}^{(i)}\}$ of the equation
$T^{n}(x)-x=0$, given by
\begin{equation}
x_{\ast}^{(i)}=
\begin{cases}
i/(2^{n}-1), & i=0,2,\dots,2^{n}-2;\\
(i+1)/(2^{n}+1), & i=1,3,\dots,2^{n}-1.
\end{cases}
\end{equation}

We emphasize that although the analytical expressions derived above are
specialized to the tent map, the proposed procedure is,\textbf{\ }in general,
suitable for the computation of joint probabilities of arbitrary unimodal
maps~\cite{collet}.

\subsection{Markov Order}

We numerically investigate the Markov order of the stochastic processes
arising from the symbolic dynamics of the tent map. Recall from
Eqs.~\eqref{eq:Markov} and~\eqref{eq:CSEmarkov} that the Markov order can be
determined as the smallest nonnegative integer $k$ for which the causation
entropy $C_{(t-k)^{-}\rightarrow t}$ vanishes. The Markov order reveals the
length of the history that carries unique information about the present
symbolic state of the system.

\begin{figure}[ptb]
\centering
\includegraphics[width=0.85\textwidth]{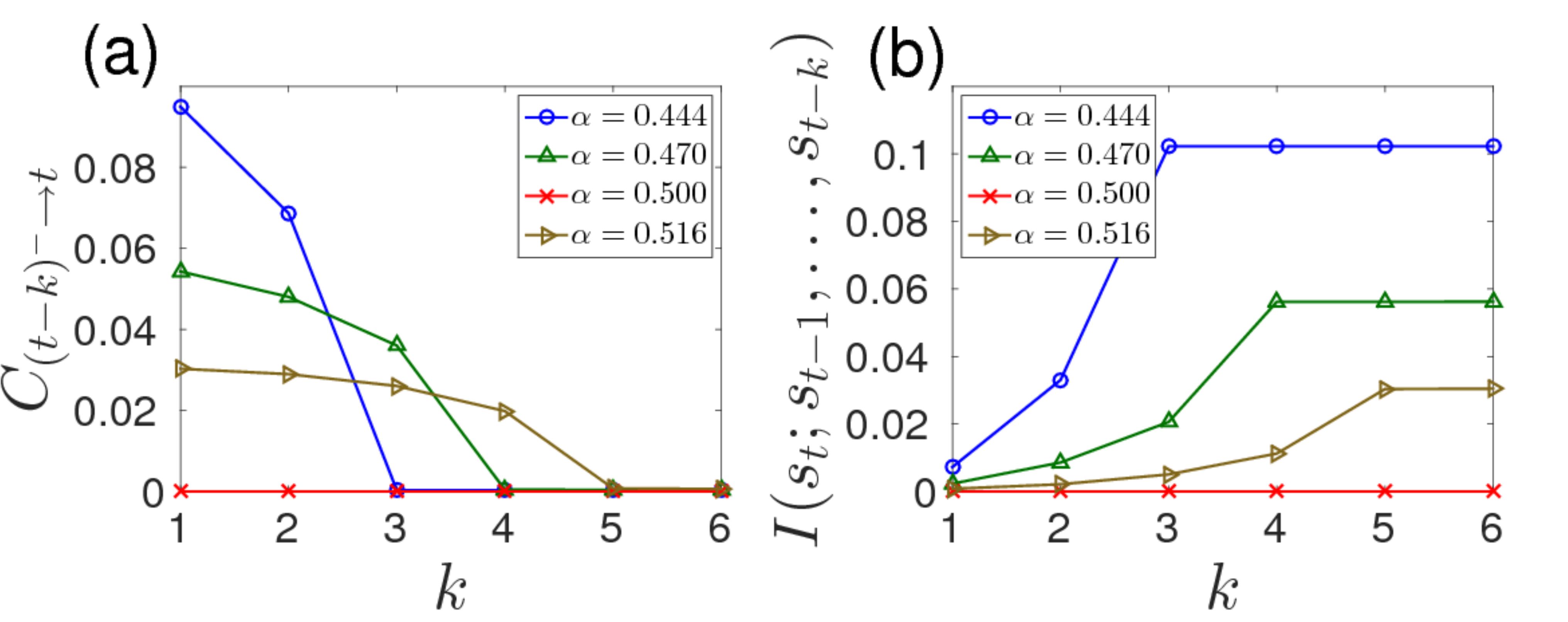}\caption{(Color online) Numerical
determination of Markov order from causation entropy. The curves show
numerically computed causation entropy $C_{(t-k)^{-}\rightarrow t}$ (a) and
mutual information $I\left(  S_{t};S_{t-1}\text{,..., }S_{t-k+1}\text{,
}S_{t-k}\right)  $(b) as functions of $k$ for various choices of $\alpha$. The
results imply that the Markov order of the symbolic dynamics of the tent map
equals $3$ ($\alpha=0.444$), $4$ ($\alpha=0.47$), $0$ ($\alpha=0.5$), and $5$
($\alpha=0.516$), respectively. In the numerical calculations, we approximate
$t^{-}$ by its finite truncation $(t-15,t-14,\dots,t-1)$.}%
\label{fig5}%
\end{figure}

Figure~\ref{fig5} shows the causation entropy $C_{(t-k)^{-}\rightarrow t}$ as
a function of $k$ for a few choices of the partitioning point $\alpha$ with
values equal to $0.444$, $0.47$, $0.5$, and $0.516$, respectively. For each
$\alpha$, the causation entropy decreases in $k$. Such monotonic dependence of
$k$ is in fact of general validity since for every $k<k^{\prime}$, the
difference of causation entropies $C_{(t-k)^{-}\rightarrow t}-C_{(t-k^{\prime
-}\rightarrow t}$ can be expressed in terms of a conditional mutual
information, which is nonnegative. On the other hand, the mutual information
$I(S_{t};S_{t-k}$, $S_{t-k+1}$,..., $S_{t-1})$ generally increases in $k$ and
saturates when the causation entropy reaches zero. Results shown in
Fig.~\ref{fig5} suggest that Markov orders can be different upon different
choices of the partition point, yielding $k=3$ for $\alpha=0.444$, $k=4$ for
$\alpha=0.47$, $k=0$ for $\alpha=0.5$, and $k=5$ for $\alpha=0.516$,
respectively. Such difference is remarkable given the relative small
differences in the values of $\alpha$.

\begin{figure}[ptb]
\centering
\includegraphics[width=0.8\textwidth]{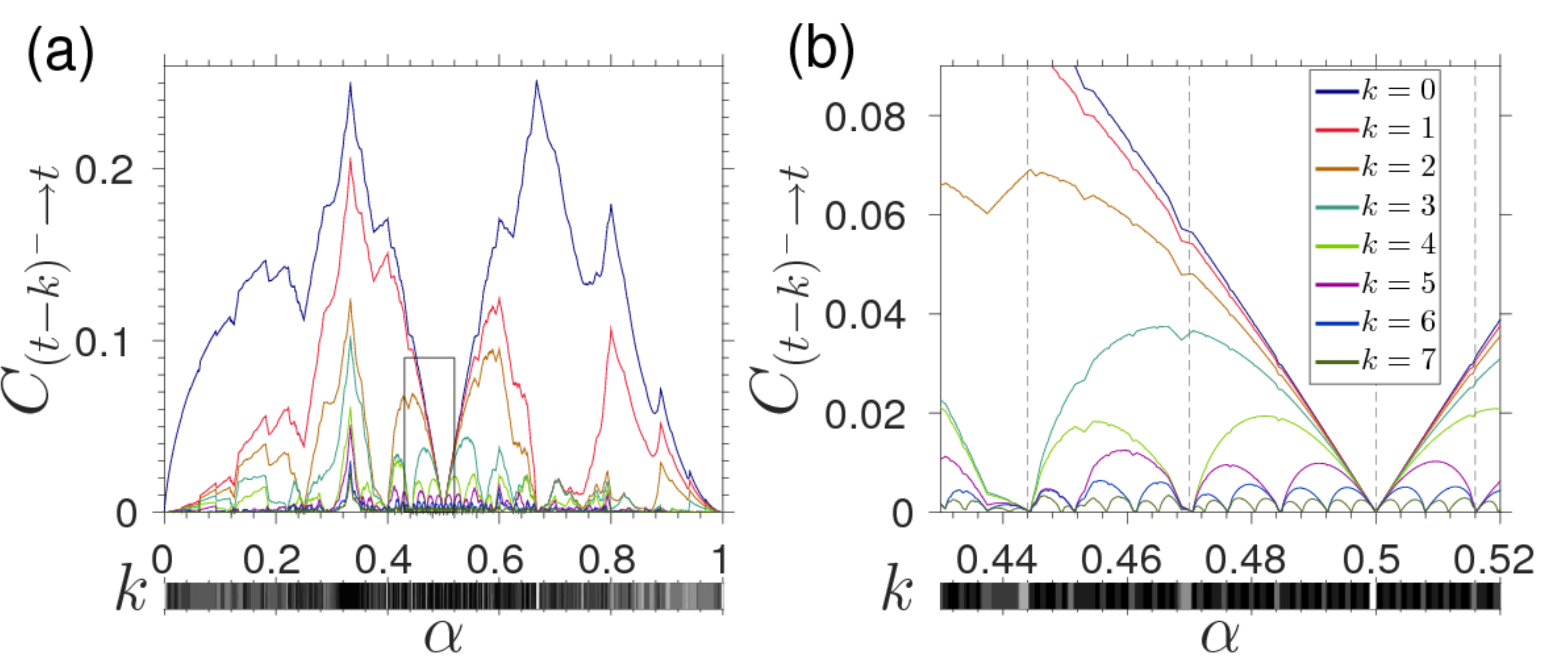}\caption{(Color online) Causation
entropies for the symbolic states of the tent map. Causation entropies
$C_{(t-k)^{-}\rightarrow t}$ (for $k=1,2,\dots,7$) are computed and shown for
a range of $\alpha$ values: $\alpha\in(0,1)$(a) and $\alpha\in(0.43,0.52)$(b).
Vertical dashed lines in panel (b) mark four specific choices of $\alpha$:
0.444, 0.47, 0.5, and 0.516, respectively. A grayscale bar is shown below each
plotting panel to visualize the numerically determined Markov order as a
function of $\alpha$, where a darker color corresponds to a higher Markov
order (white corresponds to order $0$). For each $\alpha$, the Markov order is
numerically determined as the smallest integer $k$ such that $C_{(t-k)^{-}%
\rightarrow t}<10^{-3}H(\alpha)$.}%
\label{fig6}%
\end{figure}

How does the Markov order depend on the partition point $\alpha$ in general?
We address this question by computing the causation entropy $C_{(t-k)^{-}%
\rightarrow t}$ in Eq.~\eqref{eq:CSEmarkov} as a function of $\alpha$ for a
range of $k$ values, $k=0,1,2,\dots$. The results are shown in Fig.~\ref{fig6}%
. Visually, the symbolic dynamics achieves Markov order $k$ at the values of
$\alpha$ for which all curves beyond the $(k-1)$-th one reach zero. For
example, Fig.~\ref{fig6} confirms the same Markov orders for the $\alpha$
values as shown in Fig.~\ref{fig5}. Interestingly, the Markov order seems to
depend sensitively on the choice of partitioning: a tiny bit of change in
$\alpha$ generally results in a (large) change in the Markov order. This
behavior is evident from the non-smooth and fractal appearance of the curves
in Fig.~\ref{fig6} and, from the seemingly erratic manner in which they
overlap and collapse.

\begin{figure}[ptb]
\centering
\includegraphics[width=0.8\textwidth]{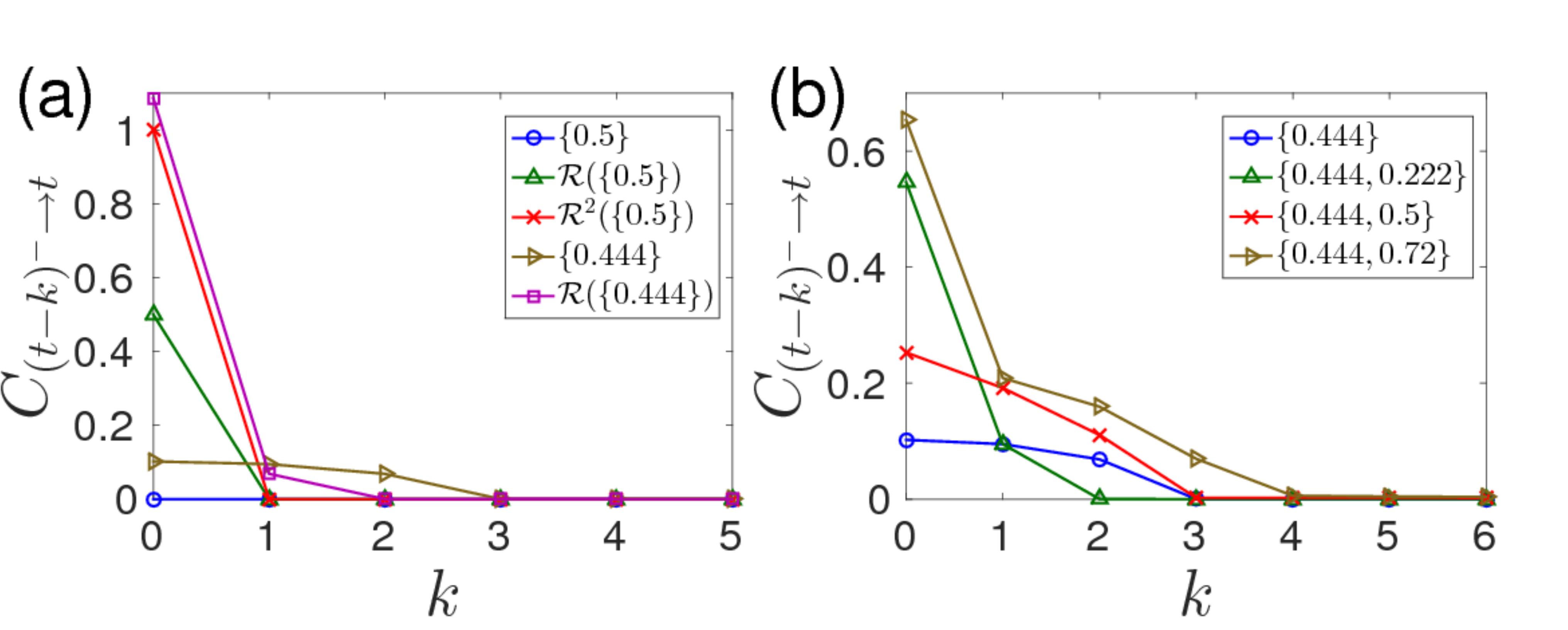}
\includegraphics[width=0.8\textwidth]{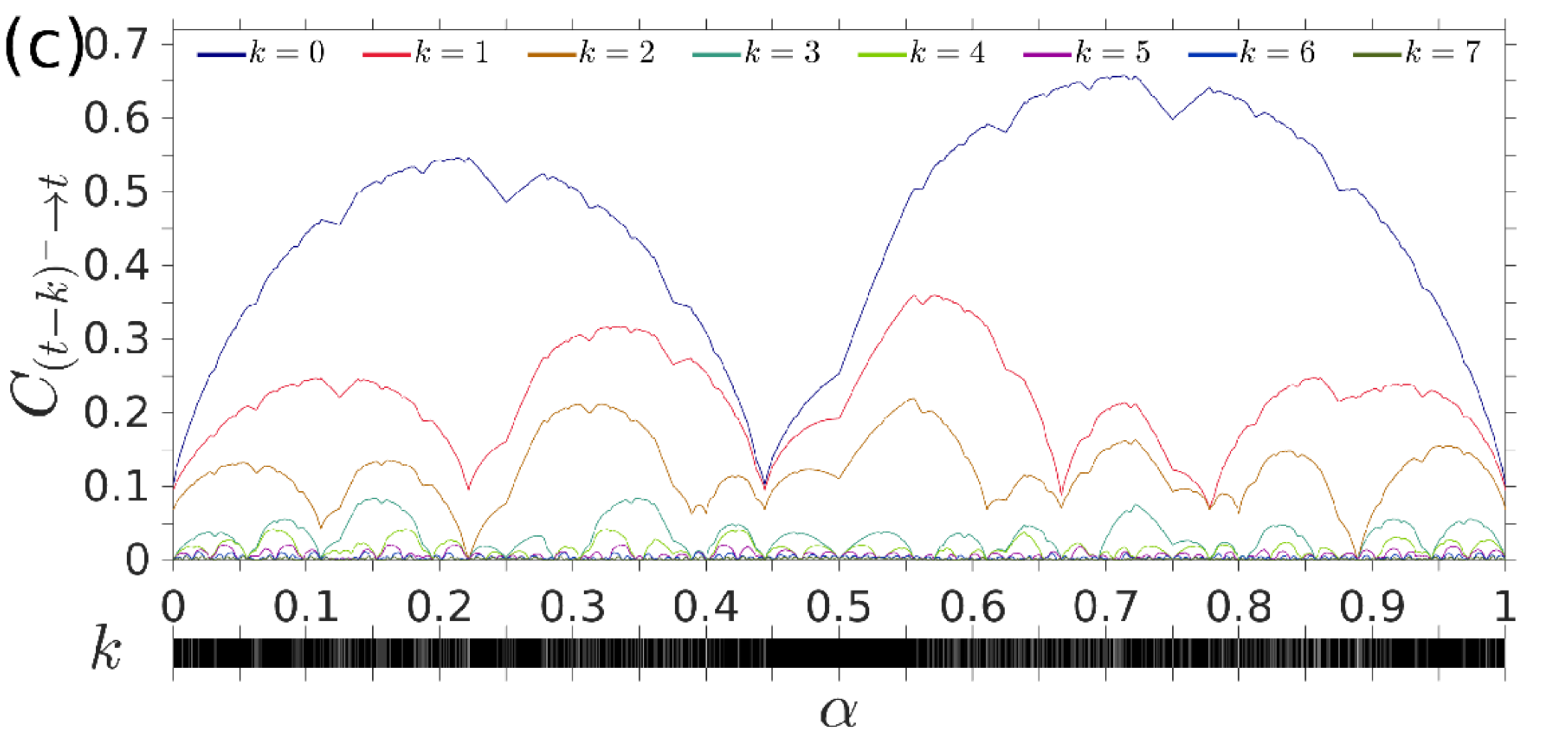}\caption{(Color online) Markov
order upon map-refinements (a) and arbitrary refinements (b)-(c). In panels
(a)-(b) the partition points are shown whereas in panel (c) the initial
partition point is fixed at $0.444 $ while the new partition point varies from
$0$ to $1$. In all calculations, we truncated $t^{-}$ as $(t-15,t-14,\dots
,t-1)$. A grayscale bar in the bottom of (c) shows the numerically computed
Markov order as a function of $\alpha$, where a darker color corresponds to a
higher Markov order (white corresponds to order $0$). For each $\alpha$, the
corresponding Markov order is computed as the smallest integer $k$ for which
$C_{(t-k)^{-}\rightarrow t}<10^{-3}H(\alpha)$.}%
\label{fig7}%
\end{figure}

Having explored the influence of the location of the partition point $\alpha$,
we ask: how do partition \textit{refinements} affect the Markov order? We now
extend our investigation to non-binary symbolic descriptions of the tent map.
Consider a \textit{map refinement} of a given partitioning $\mathcal{A}%
=\{A_{0},A_{1},\dots,A_{m}\}$~\cite{}, which is given by the intersection of
the original partition elements and their preimages under $f$, as
\begin{equation}
\mathcal{R}(\mathcal{A})\overset{\text{def}}{=}\{f^{-1}(A_{i})\cap
A_{j}\}_{i,j=0}^{m}.
\end{equation}
Inspecting Eq.~\eqref{eq:prob2} and the definition of Markov order given by
Eq.~\eqref{eq:Markov}, we conclude that if the Markov order resulting from the
original partition $\mathcal{A}$ is $k$, then the Markov order upon the
map-refinement partition $\mathcal{R}(\mathcal{A})$ equals $k-1$ if $k>1$, and
is less or equal to $1$ if $k\leq1$ (see proof in the Appendix). This result
is numerically confirmed in Fig.~\ref{fig7}(a) for the tent map. In
particular, for the original partition point $\alpha=0.5$, the Markov order
equals $0$ and map refinement increases it by $1$ while further map refinement
does not change the order. On the other hand, for $\alpha=0.444$ which yields
Markov order $3$, each map refinement decreases its order by $1$ until the
order reaches $1$. Interestingly, the same does not hold true for
\textit{arbitrary} refinements of the partition. Fig.~\ref{fig7}(b) shows that
a general refinement can either increase, decrease, or maintain the Markov
order of the resulting process. There seems to be no predicable pattern for
which the Markov order changes upon arbitrary refinement. This behavior is
further explored in Fig.~\ref{fig7}(c), which shows that for a specific
initial partition (here $\alpha=0.444$), different locations of the new
partition point generally result in different Markov orders. Once again, such
behavior appears in an irregular pattern.

\subsection{Causal Structure}

Finally, we turn to the causal structure of a symbolic dynamics, which
provides a description of the process finer than the Markov order. Unlike the
Markov order, causal structure quantifies the minimal amount of the past
history that is needed to mitigate the uncertainty about the present symbolic state.

\begin{figure}[ptb]
\centering
\includegraphics[width=0.85\textwidth]{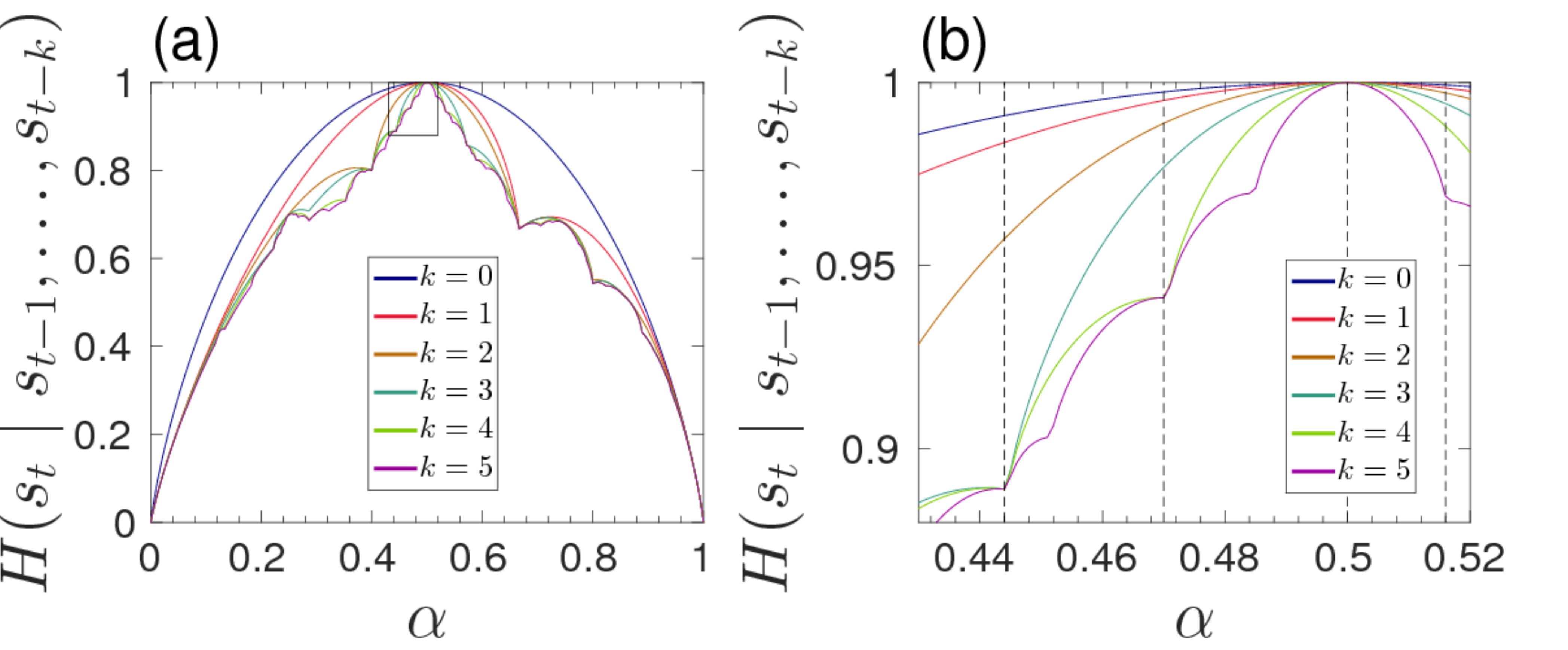}\caption{(Color online)
Uncertainty quantification of symbolic states of the tent map. (Conditional)
entropies $H\left(  S_{t}\left\vert S_{t-1}\text{,..., }S_{t-k+1}\text{,
}S_{t-k}\right.  \right)  $ for values of $k=0,1,\dots,5$, for the entire
range of $\alpha\in(0,1)$(a) and a subrange $\alpha\in(0.43,0.52)$(b).
Vertical dashed lines in both panels mark four specific choices of $\alpha$:
0.444, 0.47, 0.5, and 0.516, respectively. }%
\label{fig8}%
\end{figure}

For the tent map, the uncertainty of the symbolic state as measured by the
entropy $H(S_{t})$ achieves its maximum at $\alpha=0.5$. Including information
of past states generally reduces the uncertainty, as shown in Fig.~\ref{fig8}%
(a), except at $\alpha=0.5$, which is in fact a point for which the symbolic
dynamics is topologically conjugate (equivalent) to the original one. The fact
that the $\alpha=0.5$ partition creates an i.i.d. process is interesting
because from the dynamic equation of the system, states that are adjacent in
time are intimately linked and expected to be causally related. An important
conclusive message here is the following: partitioning of the phase space that
results in a symbolic dynamics that is equivalent to the original dynamics can
in fact yield a causal structure which differs significantly from that
inferred from the form of the equations of the original system.

\begin{figure}[ptb]
\centering
\includegraphics[width=0.9\textwidth]{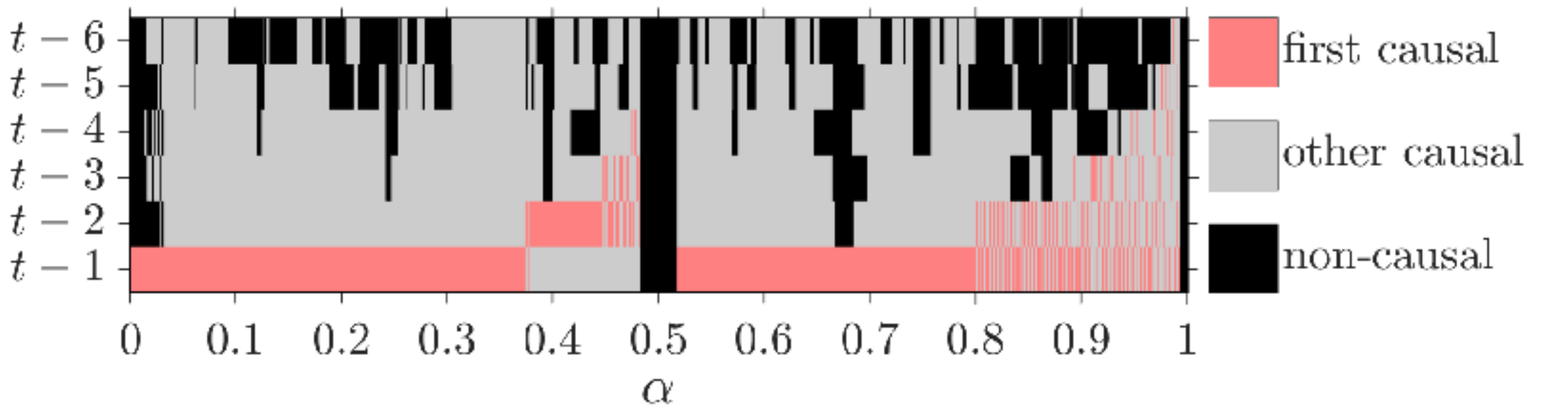}\caption{(Color online) Causal
structures from the symbolic dynamics of the tent map when the partition point
$\alpha$ is chosen from $0, 0.001, 0.002, \dots, 0.999, 1$. For each $\alpha$
we distinguish the first causal parent computed from the forward (aggregative
discovery) step of the oCSE algorithm (light red), all causal parents of $t$
from the set $\{t-1,t-2,\dots,t-6\}$ (gray), and noncausal components (black).
In all computations we used a threshold $10^{-3}H(\alpha)$ under which
causation entropy is regarded as zero. }%
\label{fig9}%
\end{figure}

Recall that a process is Markov of order $k$ if no further reduction is
possible beyond the $k$-th past state. However, the extent to which
uncertainty is reduced does not need to be monotonic in time indices. In other
words, the immediate past does not necessarily encode the most amount of
information about the present state. In fact, for several values $\alpha$
(e.g., $\alpha=0.444$ and $\alpha=0.47$), the difference between conditional
entropy $H(S_{t}|S_{t-k,\dots,t-1})$ for consecutive $k$'s is not
monotonically decreasing in $k$ [Fig.~\ref{fig8}(b)], vertical spacing between
curves). Applying the oCSE algorithms to infer the causal structure for these
$\alpha$ values, we confirmed the Markov order previously computed, and more
importantly, found that the relative importance of past time states are
ordered in a non-monotonic manner, namely $(t-2,t-3,t-1)$ for $\alpha=0.444$
(Markov order $k=3$) and $(t-3,t-2,t-1,t-4)$ for $\alpha=0.47$ (Markov order
$k=4$). We examine all values of $\alpha$ in the interval $[0,1]$ in a uniform
manner: $\{0, 0.001, 0.002, \dots, 0.999, 1\}$, using a threshold value of
$10^{-3}H(\alpha)$ for the causation entropy at the given $\alpha$. The
results are shown in Fig.~\ref{fig9}. In particular, we found several
examples for which the Markov order satisfies $k\leq6$ while the number of
causal parents is strictly less than $k$ (i.e., certain Markov time indices
are skipped in the causal structure).

\section{Summary and Final Remarks}

Symbolization is a common practice in data analysis: in the field of dynamical
systems, it bridges topological dynamics and stochastic processes through
partitioning/symbolization of the phase space; in causality inference, it
allows for the description of continuous random variables by discrete ones.
Symbolized data, in turn, are not as demanding in terms of precision and are
often considered
more robust with respect to parameters and
noise~\cite{staniek,kugi,crutchfield}.

Motivated by the problem of uncovering causal structures from finite, discrete
data, we investigated the symbolization of outputs from a simple dynamical
system, namely the tent map. We provided a full description of the joint
probabilities occurring from partitioning/symbolization of the phase space and
investigated how Markov order and causal structure can be determined from
these probabilities in terms of causation entropy, an information-theoretical
measure. We found that in general, partitioning of the phase space strongly
influences the Markov order and causal structure of the resulting stochastic
process in an irregular manner which is difficult to classify and predict. In
particular, a small change in the partition can lead to relatively large and
unexpected changes in the resulting Markov order and causal structure. To the
best of our knowledge, this is the first attempt in the literature that aims
at unravelling the intricate dependence of inferred causal structures of
dynamical systems on their different symbolic descriptions analyzed in an
information-theoretic setting. Furthermore, although the effects of map
refinements are well understood, it remains a main challenge to discover the
exact consequences of arbitrary refinements. Especially for this reason, we
have left the application of our approach to more complex dynamical systems
and/or experimental time-series data to future investigations.

On a different perspective, we note that although\textbf{\ }finding partitions
that preserve dynamical invariants (i.e., generating partitions) are known to
be a real challenge especially for\textbf{\ }high-dimensional
systems~\cite{grass1985,lai1999,davidchack2000}, it is yet unclear whether or
not such challenge remains when considering partitions that maintain Markov
order and/or causal structure. This venue of research can be especially
interesting to explore given recent advances in many different perspectives on
partitioning the phase space including adaptive binning~\cite{darbellay},
ranking and permutation of variables
\cite{bandt2002,amigo2005,staniek,kugi,haruna2013,pompe2011}, and
nearest-neighbor statistics
\cite{frenzel07,kraskov04,vejmelka08,vlachos10,porta2013}.

Finally, we remark that the non-uniqueness of symbolic descriptions of a
system implies that important concepts such as the Markov order and causal
structure are not necessarily absolute concepts: rather, they unavoidably
depend on the observational process, just like classical relativity of motion
and quantum entanglement~\cite{cafaro12}. This, in turn, suggests the
possibility of the causal structure of the very same system to be perceived
differently, even given unlimited amount of data. The concept of causality,
therefore, is observer-dependent.

\begin{acknowledgments}
We thank Dr.~Samuel~Stanton from the United States Army Research Office (ARO)
Complex Dynamics and Systems Program for his ongoing and continuous support.
This work was funded by ARO Grant No.~W911NF-12-1-0276.
\end{acknowledgments}

\appendix

\section{Monotonic Dependence of Markov Order on Map Refinements}

We will prove that for a transformation $f$ that has a uniquely ergodic
invariant probability measure $\mu$, the Markov order of the stochastic
process resulting from a partition $\mathcal{A}$ of the phase space decreases
strictly by one under a map refinement of the partition unless the original
Markov order is less or equal to one.

\noindent\textbf{Definition: Markov order of a partition.} Consider a
measure-preserving transformation $f:M\rightarrow M$ on a compact metric space
with a uniquely ergodic invariant probability measure $\mu$~\cite{KatokBook}.
Let $\mathcal{A}=\{A_{i}\}_{i=0}^{m}$ be a measurable partition of the phase
space that yields a stochastic process with time-invariant joint
probabilities
\begin{equation}
P(s_{t}=i_{t},s_{t-1}=i_{t-1},\dots,s_{t-\ell}=i_{t-\ell}) \overset
{\text{def}}{=}\mu\left(  A_{i_{t-\ell}}\cap f^{-1}(A_{i_{t-\ell+1}})\dots\cap
f^{-\ell}(A_{i_{t}})\right)  .
\end{equation}
If such a process is Markov of order $k$, we define the Markov order of the
partition to be $k$.

\noindent\textbf{Remark:} In the definition, the uniqueness of the invariant
measure implies ergodicity and ensures the well-definiteness of the joint
probabilities~\cite{KatokBook}.

\noindent\textbf{Definition: map refinement.} Consider a measure-preserving
transformation $f:M\rightarrow M$ with a probability measure $\mu$. The map
refinement of a given measurable partition $\mathcal{A}=\{A_{i}\}_{i=0}^{m}$
is defined as the partition
\begin{equation}
\mathcal{R}(\mathcal{A})\overset{\text{def}}{=}f^{-1}(\mathcal{A}%
)\vee\mathcal{A}=\{f^{-1}(A_{i})\cap A_{j}\}_{i,j=0}^{m}.
\end{equation}

\noindent\textbf{Theorem (Markov order upon map refinement.)} Consider a
measure-preserving transformation $f:M\rightarrow M$ on a compact metric space
with a uniquely ergodic invariant probability measure $\mu$. Let
$\mathcal{A}=\{A_{i}\}_{i=0}^{m}$ be a partition of $M$ and $\mathcal{R}%
(\mathcal{A})$ be its map refinement. Suppose that the Markov order of
$\mathcal{A}$ and $\mathcal{R}(\mathcal{A})$ are $k$ and $\tilde{k}$,
respectively. It follows that $\tilde{k}\leq1$ for $k\leq1$, and $\tilde
{k}=k-1$ when $k>1$.\newline\noindent\textit{Proof.} We shall denote the
probabilities resulting from the map refinement of $\mathcal{A}$ as
\begin{align}
&  \tilde{P}\left(  \tilde{s}_{t}=(i_{t},j_{t}),\tilde{s}_{t-1}=(i_{t-1}%
,j_{t-1}),\dots,\tilde{s}_{t-\ell}=(i_{t-\ell},j_{t-\ell})\right)
\nonumber\label{eq:Prefine}\\
&  \overset{\text{def}}{=}\mu\left(  \tilde{A}_{i_{t-\ell},j_{t-\ell}}\cap
f^{-1}(\tilde{A}_{i_{t-\ell+1},j_{t-\ell+1}})\dots\cap f^{-\ell}(\tilde
{A}_{i_{t},j_{t}})\right)  ,
\end{align}
where $\tilde{A}_{i,j}\overset{\text{def}}{=}f^{-1}(A_{i})\cap A_{j}$. Since
every sequence $\{\tilde{s}_{t}\}$ is determined by some orbit $\{x_{t}\}$ of
$f$ under the partition $\mathcal{R}(\mathcal{A})$, it follows that $\tilde
{s}_{t}=(i_{t},j_{t})$ if and only if $x_{t}\in f^{-1}(A_{i_{t}})\cap
A_{j_{t}}$. On the other hand, $x_{t}=f(x_{t-1})$ implies that $x_{t}\in
A_{i_{t-1}}$. Therefore $j_{t}=i_{t-1}$ in Eq.~\eqref{eq:Prefine} and
\begin{align}
&  \tilde{P}\left(  \tilde{s}_{t}=(i_{t},j_{t}),\tilde{s}_{t-1}=(i_{t-1}%
,j_{t-1}),\dots,\tilde{s}_{t-\ell+1}=(i_{t-\ell+1},j_{t-\ell+1})\right)
\nonumber\label{eq:Prelation}\\
&  =P(s_{t}=i_{t},s_{t-1}=i_{t-1},\dots,s_{t-\ell}=i_{t-\ell})
\end{align}
for all sequences $(i_{t},i_{t-1},\dots)$ with nonvanishing probability. Then,
the Theorem follows from applying Eq.~\eqref{eq:Prelation} to the definition
of Markov order given in Eq.~\eqref{eq:Markov} rewritten using the product
rule (chain rule) of conditional probability.\hfill$\blacksquare$




\end{document}